# Dielectric Bow-tie Nanocavity

Qijing Lu,[1] Fang-Jie Shu,[2] Chang-Ling Zou[1,*]

[1] *Key Laboratory of Quantum Information, University of Science and Technology of China, Hefei, Anhui 230026, China*
[2] *Department of Physics, Shangqiu Normal University, Shangqiu, He'nan 476000, China*
*Corresponding author: clzou321@ustc.edu.cn

We propose a novel dielectric bow-tie nanocavity consisting of two tip-to-tip opposite triangle semiconductor nanowires, whose end faces are coated by silver nanofilms. Based on the advantages of the dielectric slot and tip structures, and the high reflectivity from the silver mirror, light can be confined in this nanocavity with low loss. We demonstrate that the mode excited in this nanocavity has a deep subwavelength mode volume of $2.8\times10^{-4}$ μm$^3$ and a high quality factor of $4.9\times10^4$ (401.3), consequently an ultrahigh Purcell factor of $1.6\times10^7$ ($1.36\times10^5$), at 4.5 K (300 K) around the resonance wavelength of 1550 nm. This dielectric bow-tie nanocavity may find applications for integrated nanophotonic circuits, such as high-efficiency single photon source, thresholdless nanolaser, and cavity QED strong coupling experiments.

Optical microcavities [1] that confine light in very small volume can greatly enhance the light-matter interaction, so they play very important roles in various applications, including low threshold laser [2], sensor [3], nonlinear optics [4], cavity quantum electrodynamics [5] and optomechanics [6]. Miniaturization of optical microresonators is in demand to further enhance the light-matter interaction and to reduce the footprint of photonic devices for compact integrated optical circuits. However, traditional dielectric microcavities, such as whispering gallery resonators and photonic crystal cavities [1], encounter the fundamental diffraction limit of light and fail to confine light in regions smaller than half of its wavelength. Therefore, plasmonic [7-11] and metamaterial [12] nanocavities made by metal nanostructures are regarded as the most promising candidates in confining light below the diffraction limit. Unfortunately, the metal induces serious absorption loss, thus limits the quality ($Q$) factor of these nanocavities.

In this Letter, we propose a three-dimensional dielectric bow-tie (DBT) nanocavity consisting of two tip-to-tip opposite triangle semiconductor nanowires (TSNWs), which are separated by a low-index dielectric gap. Benefiting from the strong electric field enhancements at the high-index-contrast slot structure [13] and the tip of TSNWs, the proposed DBT structure confines light at truly nano-scale region (as small as $2.8\times10^{-4}$ μm$^3$ around 1550 nm), which conquers the limitation of traditional dielectric cavity. In addition, by employing metal nanofilms at the end faces of the DBT nanocavity, the $Q$ factor of this Fabry-Perot (FP) type cavity can be as large as $4.9\times10^4$ (401.3) at 4.5 K (300 K). An exceptionally high Purcell factor up to $10^7$ ($10^5$) at 4.5 K (300 K) has been predicted, which is at least one order larger than that provided by dielectric microcavities or metal nanocavities. Therefore, this DBT nanocavity holds great potential for cavity quantum electrodynamics [8], thresholdless nanolasers [14], and also will be crucial to nanoscale components in integrated photonic circuits [15]Figure 1(a) shows the schematic of the proposed DBT nanocavity. The TSNWs are identical, with a low-index dielectric gap $g$. The thickness and height of the TSNW are denoted as $t$ and $h$, respectively. The wedge tip of the TSNW has an angle of $\alpha$ and a curvature radius of $r$ (fixed to 10 nm). Light travels along the TSNWs in $zy$-direction, thus the DBT forms the FP-like nanocavity to trap the light in nanoscale volume. To reduce the reflection loss at the end faces, the top and bottom faces of the DBT nanocavity are coated by silver nanofilm. The optical properties of the DBT nanocavity are simulated by three-dimensional full-vectorial finite element method (COMSOL Multiphysics 4.3). The permittivities of the TSNW (Si) and low-index dielectric cladding (SiO$_2$) are $\varepsilon_s = 12.25$ and $\varepsilon_c = 2.25$, respectively. The permittivity of silver ($\varepsilon_m$) is set according to Johnson & Christy's experimental data [16].

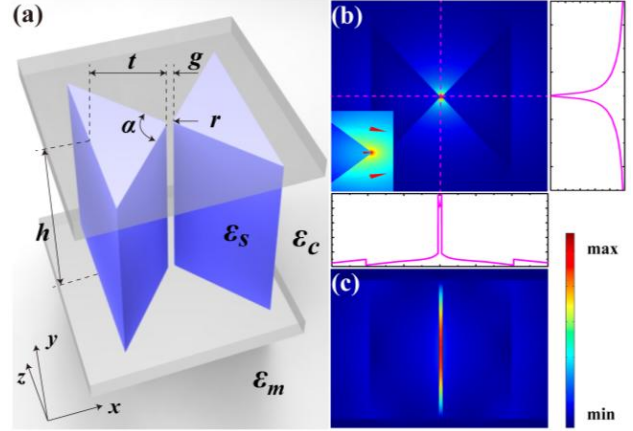

Fig. 1 (a) Schematic illustration of the bow-tie dielectric nanocavity. The origin is defined at the center of the cavity. (b) The top view ($xy$ plane) of the electric field distribution of the fundamental mode (the resonance wavelength is 1589.3 nm, $h = 400$ nm, $t = 200$ nm, $\alpha = 100°$ and $g = 10$ nm). The bottom and side panels denote the normalized electric field along the $x$ ($y = z = 0$) and $y$ ($x = z = 0$) directions, respectively. Inset: the electric field distribution around the tip of a single TSNW. The red arrows denote the direction of the electric field. (c) The side view ($xz$ plane) of the electric field distribution of the fundamental mode.

Shown in Figs. 1(b) and 1(c) are the electric field distributions of a typical fundamental TM polarized mode (electric field along x-direction) in DBT. We found that the electric field of the mode is highly squeezed into the nanometer dielectric gap, attributing to the following two mechanisms: (i) The large discontinuity of the electric field at the high-index-contrast interfaces (bottom panel of Fig. 1(b)) causes light enhancement and confinement in the slot region [13]. (ii) The local electric field enhancement in the vicinity of the wedge tip of each TSNW [17, 18], as shown by the inset of Fig. 1(b). This tip-to-tip coupling across the low-index dielectric gap enables capacitor-like energy storage, similar to that of a closely spaced dielectric nanowire and metal substrate [10, 11, 18]. We note that there are also TE polarized modes with electric field along y-direction, but with weak light confinement ability [13]. Thus, we just concern the TM polarized modes in the following studies.

We would like to emphasize that the mechanism of the extreme light confinement does not rely on the Surface Plasmon Polaritons (SPPs). Actually, the silver films are serving as mirrors to increase the end face reflectivity of this FP-type resonator. Due to the continuous boundary condition, minor of the energy (8%) inevitably penetrates to the silver, which gives rise to Ohmic loss. Our results indicate that the cavity mode volume and $Q$ factor are almost insensitive to the thickness of the silver mirror when it is greater than 40 nm.

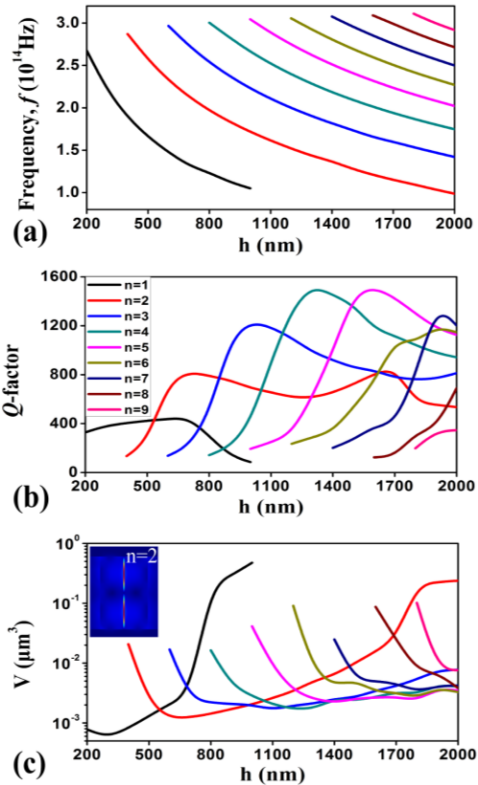

Fig. 2 The frequency (a), $Q$ factor (b) and mode volume (c) of the mode as a function of the height $h$ of the TSNM. Inset in (c): The side view ($xz$ plane) of the electric field distribution of the $n = 2$ mode.

Figure 2 shows typical results of the optical properties of the DBT nanocavity with varying $h$, which are obtained for fixed cross-sectional geometry ($t = 200$ nm, $\alpha = 100°$, $g = 10$ nm) at room temperature (300 K). There are a series of modes corresponding to different order ($n$), i.e. the number of the electric energy density antinodes along $z$-axis. According to the FP cavity model, the quantization of modes satisfies the relation

$$2hn_{eff} / \lambda \approx n, \quad (1)$$

where $n_{eff}$ is the effective mode index of the guided mode in the nanowires along z-axis. Therefore, for a certain $h$ ($n$), the mode frequency is proportional (inversely proportional) to $n$ ($h$), which agrees well with Fig. 2(a). Here, we just concern about the frequency in the range from $10^{14}$ to $3 \times 10^{14}$, because Si is lossy for higher frequency and modes are cut-off for lower frequency as silver reflectivity is very low.

Figure 2(b) plots the calculated $Q$ factor of the modes as a function of $h$, consisting of contributions from the intrinsic absorption (metal) loss and optical radiation loss ($Q^{-1}=Q^{-1}_{abs}+Q^{-1}_{rad}$). The $Q$ factors typically range from 200 to 1500 at room temperature, much higher than plasmonic and metamaterial nanocavities [7-12]. For each order $n$, there is an optimal $h$ in respect of the reflection and consequently a highest $Q$ factor. The highest $Q$ factor (~ 1500) appears for the $n = 4$ mode when $h = 1300$ nm. The upper limit for $Q$ factor results from the metal absorption loss.

Shown in Fig. 2(c) is the mode volume $V$, which is defined as the ratio of the total electric field energy density of the mode to the peak energy density [19]:

$$V = \int_{all} \varepsilon(\mathbf{r}) |E(\mathbf{r})|^2 \, d^3\mathbf{r} \Big/ \max[\varepsilon(\mathbf{r})|E(\mathbf{r})|^2]. \quad (2)$$

For each mode, $V$ first decreases then increases with increasing $h$. Mode volume smaller than $10^{-3}$ µm$^3$ can be achieved for the fundamental mode, while high Q factor still remains (> 300). The minimums of $V$ for different modes increase with $h$, because the physical cavity size is increased with $h$. When $h = 400$ nm, the $Q$ factor and $V$ of the fundamental mode are 404.4 and $8.2 \times 10^{-4}$ µm$^3$, respectively, at the resonance wavelength of 1589.3 nm.

Due to high $Q$ and ultrasmall $V$, the local density of electromagnetic states in the gap of DBT nanocavity will be significantly changed. When atoms or quantum dots are placed at the center of the DBT nanocavity, their spontaneous emission rate will be strongly modified, which is known as Purcell effect. The maximum emission rate enhancement, i.e. Purcell factor, can be expressed as

$$F_p = 3Q(\lambda/n)^3 / (4\pi^2 V). \quad (3)$$

Now, we turn to study the $F_p$ of fundamental mode in such a high $Q$ DBT nanocavity as a function of cross-sectional geometrical parameters of the TSNW, with $h$ fixed to 400 nm. The dependence on the wedge tip angle $\alpha$ when $t = 200$ nm is shown in Fig. 3(a), whereas the dependence on $t$ when $\alpha = 100°$ is shown in Fig. 3(b). The $F_p$ factors both decrease after they increase as $\alpha$ and $t$ increase, indicating the existence of optimal $\alpha$ (80°-100°) and $t$ (about 200 nm) with respect to enhancement of the spontaneous emission in the DBT nanocavity. The comparison of Figs. 3(c, e) and Figs. 3(d, f) indicates that narrower gap leads to stronger focusing of light in the gap region, similar to hybrid plasmonic waveguide [18]. When $g$ decreases from 100nm to 2 nm ($t = 200$ nm, $\alpha = 100°$ and $g = 2$ nm), the mode volume of the fundamental mode decreases significantly from $5.3 \times 10^{-3}$ µm$^3$ to $2.8 \times 10^{-4}$ µm$^3$ (orders smaller than those in other dielectric microcavities [1], comparable with that in Ref. 12) while the $Q$ factor decreases slightly from 417.0 to 401.3. It is notable that the mode volumes decrease much faster (Figs. 3(c)-(f)) than $Q$ factors when shrinking the gap width. As a result, an exceptionally high Purcell factor of $1.36 \times 10^5$, which is orders larger than those in Refs. [10-12, 19] at room temperature, can be obtained for $g$ = 2 nm (Figs. 3(a) and 3(b)).

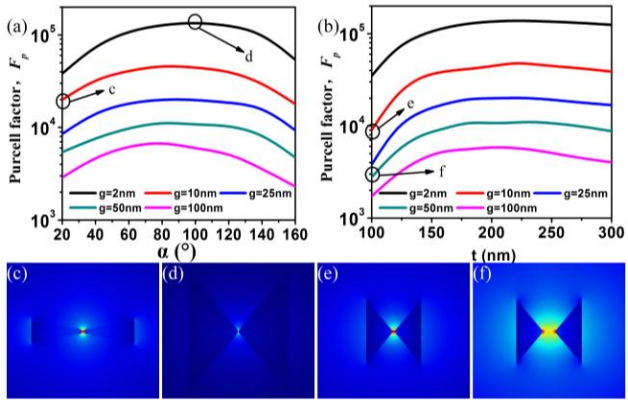

Fig. 3 (a) Purcell factor $F_p$ of the fundamental mode as a function of wedge tip angle $\alpha$ in different gap width conditions, here $h$ = 400 nm, $t$ = 200 nm. (b) $F_p$ of the fundamental mode as a function of thickness of the TSNW $t$ in different gap width conditions, here $h$ = 400 nm, $\alpha$ = 100 °. The side views ($xz$ plane) of the electric field distributions of fundamental mode for [$h$, $t$, $g$, $\alpha$] = [400 nm, 200 nm, 10 nm, 20 °] (c), [400 nm, 200 nm, 2 nm, 100 °] (d), [400 nm, 100 nm, 10 nm, 100 °] (e), [400 nm, 100 nm, 50 nm, 100 °] (f).

Although the metal absorption loss of the DBT is much lower than traditional SPP nanocavities, it still limits the Q factor greatly. For example, in the case of the strongest light confinement of the fundamental mode ($h$ = 400 nm, $\alpha$ = 100 °, $t$ = 200 nm and $g$ = 2 nm), the absorption limited Q factor is 401.33, while the radiation loss limited Q factor is $5.78 \times 10^6$. This suggests that it is instructive for DBT nanocavity operating in cryostat, where the metal absorption can be greatly suppressed in low temperature [11, 19]. As shown in Fig. 4, we investigated the dependence of the $Q$ and $F_p$ of the fundamental mode on the operating temperature (T). The Q and $F_p$ are increased by about 100 times by reducing T from room temperature to cryogenic temperature. At T = 4.5 K, both the Q factor ($4.9 \times 10^4$) and $F_p$ ($1.6 \times 10^7$) are orders larger than those of Ref. [11] at the same operating temperature

In conclusion, we proposed and numerically studied a novel dielectric bow-tie nanocavity, which consists of two semiconductor triangle nanowires posited tip-to-tip with end-faces coated by silver. We demonstrated the extreme light confinement in a nanoscale region (~$2.8 \times 10^{-4}$ μm$^3$) in this nanocavity beyond the limitation of diffraction. Another advantage of this attractive nanocavity is ultrahigh Q-factor (~$10^4$) at low temperature, since the SPP is involved. Our dielectric bow-tie nanocavity may find applications for integrated nanophotonic circuits, such as high-efficiency single photon source, thresholdless nanolaser, and cavity QED strong coupling experiments.

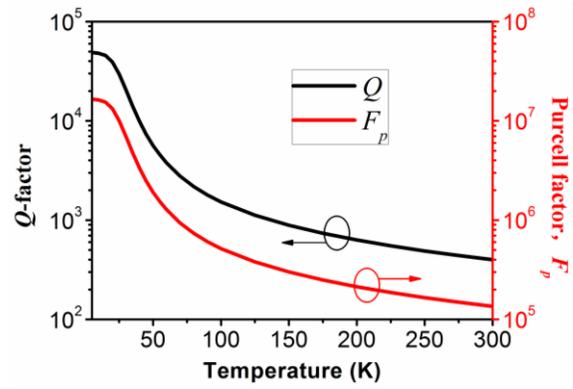

Fig. 4 (a) The $Q$ factor and Purcell factor ($F_p$) of fundamental mode as a function of operating temperatures, here $h$ = 400 nm, $t$ = 200 nm, $g$ = 2 nm and $\alpha$ = 100 °.

CLZ is supported by the 973 Programs (No. 2011CB921200). FJS is supported by the National Natural Science Foundation of China (No. 11204169 and No. 11247289).